\documentclass[preprint]{epl}
\usepackage{graphicx}
\usepackage{hyperref}
\begin{document}
\title{Boiling crisis and non-equilibrium drying transition}
\author{Vadim S. Nikolayev and Daniel A. Beysens }

\institute{D\'{e}partement de Recherche Fondamentale sur la Mati\`{e}re Condens\'{e}e, CEA Grenoble,\\ 17, rue
des Martyrs, 38054, Grenoble Cedex 9, France}
\date\today
\maketitle
\begin{abstract} Boiling crisis is the rapid formation of the
quasi-continuous vapor film between the heater and the liquid when
the heat supply exceeds a critical value. We propose a mechanism
for the boiling crisis that is based on the spreading of the dry
spot under a vapor bubble. The spreading is initiated by the vapor
recoil force, a force coming from the liquid evaporation into the
bubble. Since the evaporation intensity increases sharply near the
triple contact line, the influence of the vapor recoil can be
described as a change of the apparent contact angle. Therefore,
for the most usual case of complete wetting of the heating surface
by the liquid, the boiling crisis can be understood as a drying
transition from complete to partial wetting.
\end{abstract}


The state of nucleate boiling, which is boiling in its usual
sense, is characterized by a very large rate of heat transfer
from the heating surface to the bulk because the superheated
liquid is carried away from the heating surface by the departing
vapor bubbles. If the heating power is increased, the temperature
of the heating surface increases with the heat flux. When the
heat flux from the heater reaches a threshold value $q_{CHF}$
(the critical heat flux, CHF), the vapor bubbles suddenly form a
film which covers the heating surface and insulates the latter
from the bulk of the liquid. The temperature of the heating
surface grows so rapidly that the heater can fuse unless its
power is controlled. This phenomenon is known under the names of
``boiling crisis," ``burnout," or ``Departure from Nucleate
Boiling" (DNB) \cite{Tong}. The final state of this transition is
called film boiling.

This problem has become very important since the 1940's, with the
beginning of the industrial exploitation of heat exchangers with
large heat fluxes (as with nuclear power stations). Since then a
huge amount of research has been done for the various conditions
of pool boiling (boiling without imposed external flow) and flow
boiling (boiling of the flowing water)\cite{Katto}. Numerous
empirical correlations have been proposed, each describing the
dependence of the CHF on the physical parameters of the liquid
and of the heater more or less correctly for a particular
geometry and particular conditions of boiling \cite{Katto}. A
strong dependence of the threshold on the details of the
experimental setup coupled with difficulties in separating the
consequences of DNB from its causes is at the origin of a large
number of frequently controversial hypotheses \cite{Katto}. The
violence of boiling makes observations quite difficult. Good
quality photographic experiments are presented in only a few
articles (see e.g. \cite{Mud} -- \cite{Tor}). Despite an
increasing interest in the physical aspect of the problem during
recent years \cite{Stein,Pomo} and numerous empirical approaches,
the underlying physics still remains obscure. In this Letter, we
propose a model based on a non-equilibrium drying transition.

DNB is a really universal phenomenon which occurs inevitably for
pool as well as for flow boiling and for different flow
structures, flow velocities, liquid temperatures and pressures.
Apparently, the occurrence of the crisis does not induce any
change in the flow. The phenomenon is local \cite{Bric}: it
depends strongly only on the local values of the parameters in a
very thin layer of liquid adjacent to the heating surface. This
layer is nearly quiescent because of the no-slip boundary
condition for the fluid velocity at the heating surface. It is
thus unreasonable to assume different physical causes for DNB
according to the different conditions of pool or flow boiling,
and we think that the crisis should be induced by the {\em same
physical phenomenon}. The occurrence of DNB is influenced by the
local values of only a few parameters, the most important being
the distribution of the local temperature. As a consequence of
the local origin of DNB, the threshold depends strongly on the
wetting properties of the heating surface. {\it A priori}, it is quite
difficult to extract the dependence of the CHF on the parameters
of the heating surface because even a minor modification of its
chemical composition can cause a large change in its thermal
resistance which controls the temperature distribution and,
hence, DNB itself. However, numerous experiments \cite{Katto,wet}
show the general tendency: a poor wetting of the heating surface
by the liquid favors the DNB and vice-versa.

\begin{figure}
\centering\includegraphics[width=6cm]{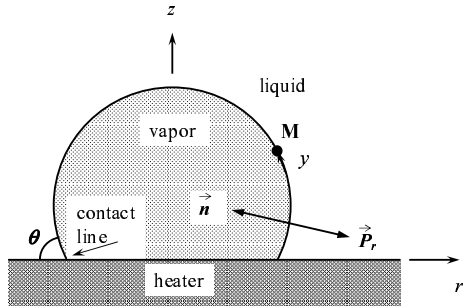} \caption{Vapor bubble on the heating surface surrounded
by liquid. The directions of the vectors $\vec{P}_r$ and $\vec{n}$ are shown as well as the axes for the
cylindrical coordinate system.} \label{Bubble}
\end{figure}

The experiments \cite{Van,Tor} in
visualization of dry spots under the vapor bubbles on the heating
surface show that at the CHF a {\em single} dry spot suddenly
begins to spread. However, its size at the CHF remains finite.
These experiments show that the bubble coalescence on the heating
surface is not the leading process. In \cite{Stein,Pav,Avk} the
vapor recoil instability \cite{Palmer} is proposed as a reason
for DNB. Although it is not clear how an instability can induce
the spreading of the dry spots, the authors show that the vapor
recoil force can be important at large evaporation rates. The
force originates in the uncompensated momentum of vapor which is
generated on the liquid-vapor interface during the
evaporation. In the reference frame of the bulk liquid, the
momentum conservation implies \begin{equation}
\vec{P}_r+\eta(\vec{v}_V+\vec{v}_i)=0, \label{Pri}
\end{equation} where $\vec{P}_r$ is the vapor recoil force per
unit interface area, $\eta$ is the evaporated mass per unit time and
unit interface area, $\vec{v}_i$ is the interface velocity, and
$\vec{v}_V$ is the vapor velocity with respect to the interface.
It is easy to establish that $\vec{v}_i=-\eta/\rho_L\;\vec{n}$,
where $\vec{n}$ is a unit vector normal to the interface directed
inside the vapor bubble (Fig.~\ref{Bubble}).
The mass conservation on the interface yields $\vec{v}_V=- \rho_L/\rho_V\;\vec{v}_i$, where $\rho_L$ and
$\rho_V$ are the mass densities of the liquid and the vapor. Therefore, (\ref{Pri}) implies\cite{Palmer}
\begin{equation}
\vec{P}_r=-\eta^2(\rho_V^{-1}-\rho_L^{-1})\vec{n}. \label{Pr}
\end{equation}
The surface deformation caused by this force is important whenever the evaporation is strong, e. g. during
high-power welding \cite{weld} or for evaporation at low pressures\cite{Palmer}.

The rate of evaporation $\eta$ can be related to the local heat flux across the interface $q_L$ by the equality
\begin{equation} q_L=H\eta, \label{eta}
\end{equation} where $H$ is the latent heat of evaporation. Hereafter, we
neglect heat conduction in the vapor with respect to the latent heat effect.

Below, we treat only the case of system at high pressure, comparable to the
critical pressure for sake of simplicity. Then
the growth of the bubble is slow enough to let the surface
tension equilibrate its shape, and the forces of hydrodynamic
origin can be neglected \cite{Tong}. This allows the problem to
be considered in the quasi-static approximation. However, the
vapor recoil remains the leading effect for any pressure.

The spreading of the dry spot looks similar to the spreading of a
liquid that wets a solid. But in the case of DNB, it is {\it vapor}
which seems to ``wet" the solid. This never happens for a non-metal
liquid under equilibrium conditions (zero heat flux) on a perfectly
clean and smooth metal surface \cite{DeG}, the finite contact angle
being possible due to the surface defects only. We show below that
a kind of drying transition occurs due to the vapor recoil force at
some heat flux that we associate with the CHF.

Using the quasi-static approximation, the variational approach
\cite{Finn} can be applied to analyze the shape of a vapor bubble
just before the boiling crisis. The free energy of the system
consists of two parts. The first part is conventional \cite{Finn}
\begin{equation} U_1=\sigma A +\sigma_{VS} A_{VS} + \sigma_{LS}
A_{LS} -\lambda V\label{U1},
\end{equation}
where $\sigma$, $\sigma_{VS}$, and $\sigma_{LS}$ are the surface tensions
for vapor-liquid, vapor-solid and liquid-solid interfaces respectively; $A_{VS}$ and
$A_{LS}$ are the corresponding interface areas; and $A$ is the area of
the vapor-liquid interface (Fig.~\ref{Bubble}). The last term in (\ref{U1}) reflects the fact that the shape of the bubble should be found for its given volume $V$, $\lambda$ being the Lagrange multiplier. The equation $\delta U_1=0$ where $\delta U_1$
is the energy change due to an infinitesimal displacement
$\delta\vec{r}$ of $A$ \cite{Finn} yields the classical conditions
of the bubble equilibrium in the absence of the external forces.

The second part $U_2$ of the free energy accounts for the virtual work of
the external forces:
\begin{equation} \delta U_2=-\int\limits_{(A)}
\vec{P}_r\cdot\delta \vec{r}\;{\rm d}A. \label{U2}
\end{equation}
The minimization $\delta U_1+\delta U_2=0$ of the total energy
leads \cite{Finn} to two equations. The first is the condition for local equilibrium of the interface
\begin{equation} K\sigma=\lambda+P_r,\label{surf}
\end{equation}
where $K$ is the local curvature of the bubble and $P_r =|\vec{P}_r|$. The second equation is $\cos\theta= c$, where $c=(\sigma_{VS}-\sigma_{LS})/\sigma$ and $\theta$ is the contact angle (Fig.~\ref{Bubble}).
For the case $c>1$ (as for the case of water on metal surface) the second equation should be substituted by the condition $\theta=0$.

Let us denote by $y$ the distance along the bubble contour
measured from the triple line to a given point M as shown in
Fig.~\ref{Bubble}. To find the bubble shape by solving Eq. (\ref{surf}) we need to know the vapor recoil as a function of $y$. In the following, we introduce  a rough approximation to solve the very complicated problem of the
heat exchange around the growing bubble. The case of saturated boiling is assumed. Thus the vapor-liquid interface is maintained at temperature $T_s$, the saturation temperature at the system pressure. We also
assume for simplicity that the thermal effect of convection can be taken into
account by renormalizing the liquid thermal conductivity. To estimate how $P_r$ varies near the contact line (i.e. when $y\rightarrow 0$) we
suppose the bubble to be two-dimensional with the contact angle $\theta=\pi/2$. Since we describe the heat exchange in a thin layer adjacent to the heating surface, we can
imagine the bubble contour $A$ to be a line $Oy$
perpendicular to the $Ox$ heater line. Then $q_L$ can be
obtained from the solution of a simple two-dimensional problem of
unsteady heat conduction in a quarter plane $x, y >0$, the point
$O(x=0, y=0)$ corresponding to the contact line.
The boundary and the initial conditions for this problem can be written in the form
$$ T_L|_{x=0}=T_s,\qquad -k_L\left.{\partial
T_L\over\partial y}\right|_{y=0}=q_S,\qquad
\left.T_L\right|_{t=0}=T_s,$$
where $T_L(x,y,t)$ is the liquid temperature, $k_L$ is the liquid thermal conductivity, and $q_S$ is the heat flux
from the heating surface which is assumed to be uniform for the
case of the thin heating wall. The solution for this problem of heat
conduction reads
\begin{equation} T_L=T_s+{q_S\over
k_L}\sqrt{\alpha_L\over\pi}\int\limits_0^t{{\rm
d}t\over\sqrt{t}}\; \mbox{erf}\left({x\over 2\sqrt{\alpha_L
t}}\right) \exp\left(-{y^2\over 4\alpha_L t}\right), \label{TL}
\end{equation}
where $\alpha_L$ is the liquid thermal diffusivity.
This solution implies the following expression for $q_L(y)$:
\begin{equation} q_L=-k_L\left.{\partial T_L\over\partial
x}\right|_{x=0}= -{q_S\over\pi}E_1\left({y^2\over 4\alpha_L
t}\right).\label{qL}
\end{equation} The exponential integral $E_1(y)$ \cite{Abr} decreases as
$\exp(-y)/y$ as $y\rightarrow\infty$. Therefore $P_r$
(related to $q_L$ via (\ref{Pr}) and (\ref{eta})) is a rapidly
decreasing function of $y$. Note that
$E_1(y)$ diverges logarithmically at the point $y=0$. The
divergence demonstrates that the evaporation is strongest
in the vicinity of the contact line. Since this singularity is integrable, the heat flux through any finite part of the interface is finite.

In the following we will use for illustration the dependence $P_r(y)$ in the form that retains the main physical features
\begin{equation}
P_r=-C\log(y/L)\exp[-(y/y_r)^2],\label{logexp}
\end{equation}
where $L$ is the length of the half-contour of the 3D axially symmetrical bubble. The coordinate $y$ is measured along the contour from the triple line as shown in Fig.~\ref{Bubble}. The characteristic length of the vapor recoil decay $y_r$ changes in time and is proportional to $\sqrt{\alpha_L t}$ (cf. (\ref{qL})). Meanwhile, the bubble grows and its radius is proportional to the same factor \cite{Tong} during the late stages of its growth. Therefore, $y_r$ is proportional to the bubble size, this fact is taken care of by the expression $y_r=aL$, where $a$ is the non-dimensional fraction of the bubble surface on which the vapor recoil is important. From the physical point of view, $y_r$ characterizes the width of the superheated layer of liquid, which is always less than the bubble size \cite{Tong}, thus $a\ll 1$. This allows the upper limit of integration to be put to infinity in the following expression for the non-dimensional strength of the vapor recoil
\begin{equation}
N_r={1\over\sigma}\int\limits_0^\infty P_r\;{\rm d}y. \label{Nr}
\end{equation}
The integration can be performed analytically yielding the relation between $C$ and $N_r$: $N_r=CaL/(4\sigma)\sqrt{\pi} [\gamma+\log (4/a^2)]$,
where $\gamma=0.577\dots$ is Euler's number \cite{Abr}.

To estimate $N_r$ at CHF when the radius of the dry spot is of the order of the radius of the drop $R$ we assume that heat is transferred to the fluid only through the belt of the width $y_r$ at the foot of bubbles. Then $q_S\sim 2\pi Ry_rn_bq_L$, where $n_b$ is a number of the bubbles per unit interface area of the heater. At CHF a large part of the heater is covered by the bubbles, we assume it to be 50\% for the estimation: $n_b\pi R^2\sim 0.5$. Then we obtain $q_L\sim q_S/a$ using $y_r\sim aR$. Since $N_r\sim q_L^2y_r/(H^2\rho_V\sigma)\sim q_S^2R/(aH^2\rho_V\sigma)$, the estimate gives $N_r\sim 1$ for $a\sim 0.01$ (usual for large bubbles, see \cite{Cooper}) and for the parameters characteristic for water at high pressures: $R\sim 1$mm, $q_S=q_{CHF}\sim 1$MW/m${}^2$, $H\sim 1$MJ/kg, $\rho_V\sim 10$kg/m${}^3$, and $\sigma\sim 10^{-2}$N/m.

Using the cylindrical $r-z$ system of coordinates (see Fig.~\ref{Bubble}), (\ref{surf}) can be written in parametric form as a system of three ordinary differential equations:
\begin{eqnarray}
{\rm d}r/{\rm d}y=\cos u,\\
{\rm d}z/{\rm d}y=\sin u,\\
{\rm d}u/{\rm d}y=-(\sin u)/r +\lambda+ P_r(y)
\end{eqnarray}
with the boundary conditions
\begin{equation}
z=0,\quad u=\theta\quad\mbox{at}\quad y=0;\qquad
r=0,\quad u=\pi\quad\mbox{at}\quad y=L.
\end{equation}
The 4th condition is necessary to determine the unknown $L$. The mathematical problem is completed by the equation, which allows the Lagrange multiplier $\lambda$ to be determined:
\begin{equation}
V=\pi\int\limits_0^L r^2\sin u \;{\rm d}y,\label{V}
\end{equation}
where $V$ is the given bubble volume. In the following, the case of the water on metal heater is considered, $\theta =0$.

\begin{figure}
\centering\includegraphics[width=8cm]{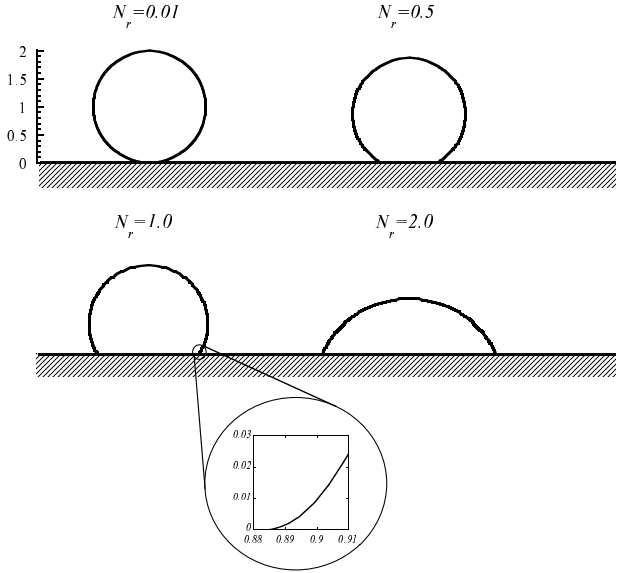} \caption{Shape of the 3D axisymmetrical vapor bubble on
the heating surface under action of the vapor recoil force calculated for $a=0.01$ (see text). The volume $V$ is
the same for all bubbles. The actual contact angle is zero as demonstrated in the insert. The coordinates are
scaled by $(3V/4\pi)^{1/3}$.}\label{Shape}
\end{figure}

The solution of the problem (\ref{logexp} -- \ref{V}) is presented in Fig.~\ref{Shape}.
Note that although the actual contact angle is zero for all of the curves (see the insert), the apparent contact
angle grows with $N_r\propto q_S^2$. At small values of $q_S$ the dry spot under the bubble corresponds to the
size of the vapor bubble nucleation spot (assumed to be zero for the calculations of Fig.~\ref{Shape}). The
bubble grows with its contact line pinned at the defects on the surface of the heater until bubble departure
under action of gravity or external hydrodynamic forces. The departure size of the bubble is small because the
adhesion (which is proportional to the contact line length \cite{Held}) is small. At some value of $q_S$ the
contact line can depin under action of the vapor recoil before the bubble detaches from the heating surface.
Fig.~\ref{Shape} shows that the dry spot reaches the size equal to the bubble diameter at $N_r\sim 1$, the value
that compares well with our estimation. The adhesion force increases with the increase of the dry spot thus
hindering the bubble departure. The departure time (the time interval during which the bubble is attached to the
heating surface) grows sharply. We think that this feedback is at the origin of the DNB, with this value of
$q_S$ to be associated with $q_{CHF}$.

There are two additional mechanisms of the dry spot growth. Each of them starts to act when the size of the dry spot attains some critical value. First, when the dry spot becomes larger than the characteristic heat diffusion length of the solid, the temperature of the dry spot increases faster than the temperature of the wetted surface. Second, as the apparent contact angle reaches $90^\circ$, the coalescence of bubbles on the heating surface increases rapidly, which causes further spreading of the dry spot. As soon as the critical size of the dry spot is reached due to vapor recoil, these two effects increase the temperature in the vicinity of the bubble thus enhancing vapor recoil and providing a feedback.

For poor wetting conditions the smaller value of the vapor recoil (and $q_{CHF}$) is needed to create a large dry spot. This explains the decrease of $q_{CHF}$ with the increase of the contact angle and surface roughness \cite{wet}.

In this Letter we have proposed a mechanism for the boiling crisis
based on a transition from complete to partial wetting of the
heating surface as due to a vapor recoil force. Unfortunately, it is quite difficult to obtain the analytical form of the DNB criterion only from the above considerations. They can be helpful, however, for future numerical simulations that take into account more realistic temperature distributions, the balance between adhesion and lift-off forces, and the changing bubble shape. This model makes the departure time be the crucial parameter that should increase rapidly near the crisis. Unfortunately, such experimental studies are not available.

It is clear now why vapor recoil can be the triggering phenomenon for the boiling crisis under various conditions of boiling. Providing that the vapor recoil force operates in a very thin belt at the base of the bubble, the change in the bubble shape can be accounted for by a change in an apparent contact angle controlled by the vapor recoil force. Another interesting result is the possibility to change the size of the dry spot under the bubble while the actual contact angle remains zero.

\acknowledgments

One of the authors (V.~N.) would like to thank the Direction of
DRFMC, CEA/Grenoble for its kind hospitality, and CEA and EDF for
financial support. The authors are grateful to Etienne Bri\`ere,
St\'ephane Pujet, Jean-Marc Delhaye et Dominique Grand for
fruitful discussions and exhaustive information on the boiling
crisis.

\end{document}